# PROBING THE LOW-LUMINOSITY XLF IN NORMAL ELLIPTICAL GALAXIES

D.-W. Kim[1], G. Fabbiano[1], V. Kalogera[2], A. R. King[3], S. Pellegrini[4], G. Trinchieri[5], S. E. Zepf[6], A. Zezas[1], L. Angelini[7], R. L. Davies[8], J. S. Gallagher[9]

[1] Harvard-Smithsonian Center for Astrophysics, 60 Garden St., Cambridge MA 02138; kim@cfa.harvard.edu, gfabbiano@cfa.harvard.edu, azezas@cfa.harvard.edu
[2] Northwestern University, Department of Physics and Astronomy, 2145 Sheridan Road, Evanston, IL 60208; vicky@northwestern.edu
[3] University of Leicester, Leicester, LE1 7RH, UK; ark@star.le.ac.uk
[4] Dipartimento di Astronomia, Universita' di Bologna, Via Ranzani 1, 40127, Bologna, Italy; silvia.pellegrini@unibo.it
[5] INAF-Osservatorio Astronomico di Brera, via Brera 28, 20121 Milano, Italy; ginevra.trinchieri@brera.inaf.it
[6] Department of Physics and Astronomy, Michigan State University, East Lansing, MI 48824-2320; zepf@pa.msu.edu
[7] Laboratory for High Energy Astrophysics, NASA Goddard Space Flight Center, Code 660, Greenbelt, MD 20771; angelini@davide.gsfc.nasa.gov
[8] Denys Wilkinson Building, University of Oxford, Keble Road, Oxford; rld@astro.ox.ac.uk
[9] Astronomy Department, University of Wisconsin, 475 North Charter Street, Madison, WI 53706; jsg@astro.wisc.edu

(May 31, 2006)

## Abstract

We present the first low luminosity ($L_X > 5 - 10 \times 10^{36}$ erg s$^{-1}$) X-ray luminosity functions (XLFs) of low-mass X-ray binaries (LMXBs) determined for two typical old elliptical galaxies, NGC 3379 and NGC 4278. Because both galaxies contain little diffuse emission from hot ISM and no recent significant star formation (hence no high-mass X-ray binary contamination), they provide two of the best homogeneous sample of LMXBs. With 110 and 140 ks *Chandra* ACIS S3 exposures, we detect 59 and 112 LMXBs within the $D_{25}$ ellipse of NGC 3379 and NGC 4278, respectively. The resulting XLFs are well represented by a single power-law with a slope (in a differential form) of 1.9 ± 0.1. In NGC 4278, we can exclude the break at $L_X \sim 5 \times 10^{37}$ erg s$^{-1}$ that was recently suggested to be a general feature of LMXB XLFs. In NGC 3379 instead we find a localized excess over the power law XLF at $\sim 4 \times 10^{37}$ erg s$^{-1}$, but with a marginal significance of $\sim 1.6\sigma$. Because of the small number of luminous sources, we cannot constrain the high luminosity break (at $5 \times 10^{38}$ erg s$^{-1}$) found in a large sample of early type galaxies. While the optical luminosities of the two galaxies are similar, their integrated LMXB X-ray luminosities differ by a factor of 4, consistent with the relation between the X-ray to optical luminosity ratio and the globular cluster specific frequency.

Subject headings: galaxies: elliptical and lenticular – X-rays: binaries – X-rays: galaxies

1. INTRODUCTION

Low-mass X-ray binaries (LMXBs) are luminous X-ray sources associated with old stellar populations, powered by the accretion of the atmosphere of a low-mass late-type star onto a compact stellar remnant, either a neutron star or black hole. The formation and evolution of LMXBs has been debated since their discovery in the Galaxy. Proposed evolutionary paths include both native field binary systems and binary formation by dynamical interactions in either the inner bulge or globular clusters (e.g., Grindlay 1984; see review in Verbunt & van den Heuvel 1995; Bildsten & Deloye 2004). This debate has been rekindled by the detection of populations of these sources in early type galaxies with *Chandra*. In these galaxies, a significant fraction of the LMXBs have been found in globular clusters, but there is also some evidence pointing to native field binaries (e.g., Maccarone, Kundu & Zepf 2003; Irwin 2005; Kim E., et al 2006 and references to earlier work therein; see review by Fabbiano 2006).

The X-ray Luminosity Function (XLF) of LMXB populations has emerged as a powerful tool to constrain the nature and evolution of these sources. The high luminosity end ($L_X >$ several $10^{37}$ erg s$^{-1}$) of the XLF is now well constrained from the study of many E and S0 galaxies observed with *Chandra* (e.g., Kim & Fabbiano 2004, Gilfanov 2004). The normalization (i.e., the total number of LMXBs in a given galaxy) is strongly related to the stellar mass of the galaxy (Gilfanov 2004; see also Kim & Fabbiano 2004), as it would be expected in the case of slowly evolving LMXBs, although a link to the specific frequency of GCs has also been reported (Kundu et al. 2002; Kim & Fabbiano 2004; Kim, E. et al. 2006). A break in this XLF (first suggested by Sarazin et al. 2001, and conclusively reported by Kim & Fabbiano 2004 and Gilfanov 2004 at $L_X = 5 \pm 1.6 \times 10^{38}$ erg s$^{-1}$) is consistent with the Eddington luminosity of either a massive neutron star (~3 $M_\odot$) or *He*-enriched accretion and may reflect the presence of both neutron star and black-hole LMXBs in the X-ray source populations (Ivanova & Kalogera 2006). This break is also predicted in the model of short-lived, high-birth-rate, ultra-compact binary evolution in GCs by Bildsten & Deloye (2004).

What has been less well constrained so far is the behavior of the XLF at the lower luminosities, which are more typical of the majority of LMXBs in the bulge of the Milky Way and M31. With the exception of NGC5128 (the radio galaxy Centaurus A), for which the XLF has been measured down to ~2 × $10^{36}$ erg s$^{-1}$ (Kraft et al 2001; Voss & Gilfanov 2005), the available *Chandra* data so far have not allowed the detection of LMXBs in E and S0 galaxies at luminosities below the mid $10^{37}$ erg s$^{-1}$ range. By including Cen A and the LMXB (bulge) population of nearby spirals in a co-added XLF, Gilfanov (2004) suggested a significant flattening of the XLF at $L_X < 5 \times 10^{37}$ erg s$^{-1}$ (see also Voss & Gilfanov 2005). This flattening, first reported in the LMXB XLF of the Milky Way (Grimm, Gilfanov & Sunyaev 2002) is suggested in a number of models, albeit because of different mechanisms (e.g., Bildsten & Deloye 2004; Pfahl, Rappaport & Podsiadlowski 2003; Postnov & Kunanov 2005). The 'outburst peak luminosity – orbital period' correlation (King & Ritter 1998) predicts a break at this luminosity if a large fraction of the sources are short-period (~ 1 hour) binaries. Similarly, ultra-compact (~10 minute period) LMXBs formed in GCs would show a break at ~$10^{37}$ erg s$^{-1}$ in absence of X-ray heating, while with X-ray heating the disk can stay stable to lower luminosities, so no XLF break would occur (Bildsten & Deloye 2004). It is therefore important to study the low-luminosity XLF of normal elliptical LMXB populations to confirm the presence and ubiquity of this break. Moreover, detailed studies of the globular cluster LMXB population of M31 have reported a distinctive break at ~$2\times10^{37}$ erg s$^{-1}$ (Kong *et al.* 2003; Trudolyubov & Priedhorsky 2004). The discovery of a similar break in the E and S0 XLFs may argue for a GC-LMXB connection in these galaxies.

In this paper we report the first direct measurement of the low-luminosity XLF in normal nearby elliptical galaxies. The two galaxies NGC3379 and NGC 4278 (Table 1) are being observed as part of a *Chandra* legacy program to study in depth the LMXB population. Because both elliptical galaxies are old (e.g., Trager et al. 2000; Terlevich & Forbes 2002), they provide a clean sample of LMXBs with no contamination by HMXBs which are likely to contribute to the X-ray source populations of spiral galaxies (the Milky Way and M31) and young or rejuvenated E and S0 galaxies resulting from recent mergers (e.g., NGC 5128). While we expect to reach a limiting luminosity of a few $10^{36}$ erg s$^{-1}$ at the end of this observing campaign, our first observations, in conjunction with the data in the *Chandra* archive, result in total exposures of 110 and 140 ks, respectively. The corresponding limiting luminosities, restricting ourselves to data not affected by incompleteness, are 1 and 3 x $10^{37}$ erg s$^{-1}$ (for NGC3379 and NGC 4278, respectively); applying incompleteness corrections to the data, the resulting XLFs reach down to ~5 x $10^{36}$ and ~1 x $10^{37}$ erg s$^{-1}$ respectively. As discussed in this paper, we find that the conclusion of a ubiquitous flattening of the XLF is not supported by our results.

We adopt distances of 10.6 Mpc (NGC 3379) and 16.1 Mpc (NGC 4278) throughout this paper, based on the surface brightness fluctuation analysis by Tonry et al. (2001). At these distances, 1′ corresponds to 3.1 kpc and 4.7 kpc, respectively.

2. *Chandra* X-RAY OBSERVATION

NGC 3379 was observed for 85 ks on Jan. 23, 2006 with the S3 (back-illuminated) chip of *Chandra* Advanced CCD Imaging Spectrometer (ACIS) (obsid = 7073). The ACIS data were reduced in a similar manner as described in Kim & Fabbiano (2003) with a custom-made pipeline (XPIPE), specifically developed for the *Chandra* Multi-wavelength Project (ChaMP; Kim et al. 2004). Removal of background flares reduced the effective exposure time of CCD S3 to 80 ks. NGC 3379 had been previously observed with *Chandra* ACIS (obsid = 1587) for 29 ks on Feb. 13, 2001 (David et al. 2005). We have retrieved these data from the *Chandra* archive (http://asc.harvard.edu/ cda/) and reduced them with the same method after correcting instrumental effects (e.g., time-dependent QE variation) with up-to-date calibration data (http://asc.harvard.edu/ caldb/). Combining the two observations by re-projecting to a common tangent point (using ***merge_all*** available in the CIAO contributed package; see http://cxc .harvard.edu/ ciao/ threads/ combine/), we reach an effective exposure on CCD S3 of 110 ks. We show the merged image in Figure 1a, where the point X-ray sources and the optical size ($D_{25}$) are marked.

NGC 4278 was observed for 110 ks on Mar. 16, 2006 with the S3 chip of ACIS (obsid = 7077). Removal of background flares reduced the effective exposure time to 108 ks. NGC 4278 had been previously observed with ACIS (obsid = 4741) for 37 ks on Feb. 3, 2005 (PI: J. Irwin). We have also combined the two observations and the merged data correspond to an effective exposure on CCD S3 of 145 ks. The merged image is shown in Figure 1b.

```
                                   Table 1
------------------------------------------------------------------------
  name          D         D25        PA      B_T_0    M_B      L_B         S_N
              (Mpc)       (')       (deg)    (mag)   (mag)    (L_Bo)
   (1)         (2)        (3)        (4)      (5)     (6)      (7)         (8)
------------------------------------------------------------------------
NGC 3379      10.6      5.4x4.8     67.5     10.18   19.94    1.35e10      1.2
NGC 4278      16.1      4.1x3.8     27.5     10.97   20.06    1.63e10      6.9
------------------------------------------------------------------------
1. Galaxy name
2. Distance from Tonry et al. (2001)
3. Optical size (diameter) of galaxy determined at 25th magnitude from RC3
4. Position angle of the major axis from NED
5. B_T_0 from RC3
6. Absolute blue magnitude
7. Blue luminosity calculated by adopting absolute solar blue magnitude of 5.47 mag
8. Globular cluster specific frequency from Ashman & Zepf (1998).
```

## 3. X-RAY LUMINOSITY FUNCTIONS

From the merged images of NGC 3379 and NGC 4278 observations, we detect 109 and 197 point sources in CCD S3 only (column 5 in Table 2). A number of these sources vary and some are detected only in one observation. Source variability and spectral parameters will be presented in a future paper. In this paper, we only consider the effect of source variability on the observed XLF. To measure the X-ray luminosity (in 0.3-8 keV) from the merged data, we take into account the temporal QE variation ( http://cxc.harvard.edu /cal/Acis/Cal_prods/qeDeg/) by calculating the energy conversion factor (ECF) in each observation and then taking an exposure-weighted mean ECF. The ECF (0.3-8 keV) varies by ~12% between two observations of NGC 3379 and by 0.1% between two observations of NGC 4278.

To construct the XLF, we use point sources detected within the $D_{25}$ ellipse (the size and position angle are given in Table 1). Although some X-ray sources outside the $D_{25}$ ellipse may be associated with the galaxy, we exclude them to reduce the contamination by foreground or background X-ray sources. Based on the ChaMP+CDF log(N)-log(S) (Kim, M. et al. 2006), we estimate the cosmic background sources (mostly AGN) to be 4-13 (column 10 in Table 2) within the $D_{25}$ ellipse at the flux limit of 90% completeness (column 8 in Table 2 – see below for details). This results in a source contamination of 20% in NGC 3379 and 7% in NGC 4278 (see also Figure 2); these fractions would be considerably higher if we included the sources detected outside the $D_{25}$ ellipse, since they scale with the search area. We also exclude sources located near the galactic centers (R < 10″), because of the large photometric error caused by confusion with other overlapping sources and by the presence of diffuse emission; these conditions make the incompleteness corrections uncertain. In a previous work with 14 elliptical galaxies, Kim & Fabbiano (2004) excluded sources within R < 20″ from the XLF; the smaller amount of hot ISM present in both NGC 3379 and NGC 4278 allows us to use sources nearer to the galaxy centers in this case. With these selection criteria, we obtain 59 and 112 sources in NGC 3379 and NGC 4278, respectively, to build the XLFs.

To determine the XLFs accurately, it is most critical to correct for incompleteness (see Kim & Fabbiano 2003, 2004). Without such correction, the XLF would look flattened at the lower luminosities where the detection is not complete, causing an artificial break. Following Kim and Fabbiano (2004), we performed extensive simulations to generate incompleteness corrections: we simulated 20,000 point sources using **MARX** (http://space.mit.edu/ ASC/ MARX/), added them one by one to the observed image and then determined whether the added source is detected. Since we used the real observed data as the baseline, we could correct simultaneously three biases: detection limit, Eddington bias (Eddington 1913) and source confusion (Kim & Fabbiano 2003). In the simulations, we assumed a typical LMXB XLF differential slope of β=2 (Kim and Fabbiano 2004),

$$\frac{dN}{dL_X} = k L_X^{-\beta}.$$

We note that the adopted XLF slope does not significantly affect the results, because the correction is determined by the ratio of the number of input sources to that of detected sources (see also Kim and Fabbiano 2003). As shown in Kim, E. et al. (2006), the radial distribution of LMXBs closely follows that of the optical halo light, regardless of their association with globular clusters. Therefore, we adopted an $r^{-1/4}$ law for the radial distribution of the LMXBs. Also, we did not use LMXBs in the central regions, where the radial profile dependency is most significant.

We find that the 90% completeness limit (i.e., where 10% of sources with this luminosity would not be detected inside the $D_{25}$ ellipse, but excluding the central 10″) is $L_X = 1 \times 10^{37}$ erg s$^{-1}$ for NGC 3379 and $L_X = 3 \times 10^{37}$ erg s$^{-1}$ for NGC 4278 (column 9 in Table 2); we can reliably correct the XLFs to X-ray luminosities at least a factor of 2-3 lower than the 90% limit (roughly corresponding to a 50% limit).

We show the bias-corrected cumulative XLFs in Figure 2, where the corrected data are identified by squares with error bars and the uncorrected raw data by a smaller circle. It is clear that the apparent XLF flattenings at the low luminosities disappear in both cases. Using **sherpa** (http://cxc.harvard.edu/sherpa), we fit the bias-corrected XLFs (in a differential form) with single power-laws using Cash statistics (for an unbinned distribution, but read in **sherpa** as a histogram mode) and $\chi^2$ statistics (for a binned distribution). The Cash statistic (also C-stat) utilizes a maximum likelihood function and can be applied regardless of the number in each bin. For the $\chi^2$ statistics, we select the minimum to be 10 sources in each bin and apply the Gehrels variance function for the error calculation (Gehrels 1986). Both statistics result in consistent parameters within the error. The solid lines in Figure 2 indicate the best-fit single power law models determined with the Cash statistics. We list the best-fit parameters and the corresponding errors (90% confidence level) in Table 2. As indicated in Figure 2, the quality of fitting with a single power law is good. The $\chi^2$ method results in $\chi^2_{red}$ = 1.19 with 4 degrees of freedom (dof) and $\chi^2_{red}$ = 0.70 with 10 dof for NGC 3379 and NGC 4278, respectively. To check the possible XLF flattening at low luminosities, we applied a broken power law with the break luminosity varying freely. In both galaxies a slightly improved total $\chi^2$ accompanied by a lower dof resulted in a worse $\chi^2_{red}$, indicating that the broken power law is not statistically required with the given luminosity range. We also fitted the data excluding the faint sources below the 50% completeness limit, reaching consistent results within the statistical error. A persistent flattening of the XLFs at the low luminosities is clearly absent in both galaxies. However, in NGC 3379 there is a marginal (~1.6$\sigma$) excess over the best-fit power-law at $L_X$ = 3-6 x $10^{37}$ erg s$^{-1}$.

Table 2

| name | obsid | obs date | net exposure (ks) | # src S3 | # src R25 | # src r>10" | 90% compleness limit Fx | 90% compleness limit Lx | # bkg src | slope (err) | amp (err) |
|---|---|---|---|---|---|---|---|---|---|---|---|
| (1) | (2) | (3) | (4) | (5) | (6) | (7) | (8) | (9) | (10) | (11) | (12) |
| NGC 3379 | 01587 | Feb 13, 2001 | 28.8 | 72 | 44 | 34 | 1.5(-15) | 2.1(37) | 8 | 1.8 (0.3) | 7.5 (2.6) |
| NGC 3379 | 07073 | Jan 23, 2006 | 80.3 | 86 | 57 | 45 | 0.9(-15) | 1.3(37) | 12 | 1.8 (0.2) | 6.2 (2.2) |
| NGC 3379 | merge | | 109.9 | 109 | 70 | 59 | 0.8(-15) | 1.0(37) | 13 | 1.9 (0.2) | 5.6 (1.8) |
| NGC 4278 | 04741 | Feb 3, 2005 | 36.6 | 92 | 58 | 54 | 2.2(-15) | 6.5(37) | 4 | 1.9 (0.2) | 22.9 (4.1) |
| NGC 4278 | 07077 | Mar 16, 2006 | 107.6 | 175 | 116 | 108 | 1.1(-15) | 3.3(37) | 7 | 2.0 (0.1) | 23.2 (4.6) |
| NGC 4278 | merge | | 144.6 | 197 | 122 | 112 | 0.9(-15) | 2.8(37) | 8 | 1.9 (0.1) | 23.1 (4.4) |

1. galaxy name
2. *Chandra* observation id
3. *Chandra* observation date
4. Net exposure after removing the background
5. Number of point sources detected in CCD S3
6. Number of point sources detected within the D25 ellipse
7. same as (6), but exclude sources at R<10". These sources are actually used in building XLF.
8. 90% compleness limit in the source flux (0.3-8 keV), i.e.,
   10% of sources with this flux would not be detected inside the $D_{25}$ ellipse, but excluding the central 10".
9. same as (8) but in the source luminosity (0.3-8 keV)
10. Number of cosmic background sources within the D25 ellipse at the flux limit given in column (8)
11. Best-fit XLF slope (in a differential form) and error (90% confidence)
12. Best-fit XLF amplitude and error (90% confidence)

Also plotted in Figure 2 is the expected number (dotted line) of cosmic X-ray background sources (Kim, M. et al. 2006). Because background sources are not fully detected at low fluxes (the same detection incompleteness as for the sources in the galaxies would apply), the dotted line will be flattened as the observed XLF, therefore the dotted line should be compared with the corrected XLF. It is then clear that the contamination is only a small fraction (20% in NGC 3379 and 7% in NGC 4278) and the steepening at low $L_X$ is not due to background source contamination. We repeated the XLF fit with an additional power-law component for the expected background sources and obtained consistent best-fit parameters to those listed in Table 2.

To investigate the effect of source variability on the XLFs, we compare XLFs built from the individual observations (as well as the merged observations) in Figure 3. The black and green histograms indicate the corrected and uncorrected XLFs, respectively. The vertical bars mark the 90% completeness limits (column 9 in Table 2). The XLFs from the two individual observations are shifted downward vertically by $\Delta \log(N) = -1.0$ (the shallower one) and $-0.5$ (the deeper one) for visibility. Error bars are not plotted here to avoid over-crowding, but can be seen in Figure 2 for the merged XLFs. The best-fit single power-law models (dashed blue diagonal lines) determined from the merged observations are plotted with the XLFs (both individual and merged). In general, the XLFs (both in slope and amplitude) are consistent with each other, within the statistical error, even though there are a number of variable X-ray sources. A similar result of the robust XLF in the multiple observations of the Antennae was previously reported (REF – pepi, please add this).

The validity of our low-luminosity bias correction is also verified by comparing the XLFs made from observation of different depths. The corrected XLFs from the shallower observations closely follow the XLFs from the deeper observations at the low luminosities. This is most clearly seen in NGC 4278, which hosts a larger number of LMXBs, hence has a higher statistical significance (Figure 3b). The predicted (corrected) XLF below $L_X < 6.5 \times 10^{37}$ erg s$^{-1}$ (the 90% confidence limit) of the 37 ksec observation (obsid = 04741) is well reproduced in the observed (uncorrected) XLF of 108 ks observation (obsid = 07077) down to $L_X < 3 \times 10^{37}$ erg s$^{-1}$.

The high luminosity shape of the XLF of the LMXB populations of elliptical galaxies is well established with a large sample of elliptical galaxies and can be fitted with a broken power-law model, with a break (at $5 \times 10^{38}$ erg s$^{-1}$; Kim & Fabbiano 2004, Gilfanov 2004), which may stem from the presence of NS and BH binary populations. In the current sample, we cannot constrain the high luminosity break because of a small number of luminous sources (only 2 sources in each galaxy with $L_X > 5 \times 10^{38}$ erg s$^{-1}$). We also note that in NGC 4278 (Fig 3b), the XLF of the longer observation (obsid 7077) looks steeper in the high $L_X$ range than that of the shorter observation (obsid 4741), but this is simply because the most luminous source ($L_X = 2 \times 10^{39}$ erg s$^{-1}$), seen in the shorter (earlier) observation, disappears in the longer observation taken a year later.

4. DISCUSSION

In this paper, for the first time we explore the low-luminosity XLFs of normal elliptical galaxies. This low-luminosity XLF has been studied in the Galaxy, where a flattening at the low-luminosities has been reported (Grimm, Gilfanov & Sunyaev 2002), and in M31, where diverse and complex behavior can be seen looking at different stellar fields (e.g., Kong et al. 2002, 2003; Trudolyubov & Priedhorsky 2004; see review in Fabbiano & White 2006). Gilfanov (2004) suggested that a low-luminosity flattening of the LMXB XLF is a general feature of LMXB populations. This flattening is also observed in the X-ray source population of NGC 5128 (Cen A), below $L_X = 5 \times 10^{37}$ erg s$^{-1}$ (Voss & Gilfanov 2005).

Our results suggest that the conclusion of a ubiquitous flattening of the low-luminosity shape of the LMXB XLF may be premature. In two normal elliptical galaxies,

NGC 3379 and NGC 4278, we found that the XLFs can be fitted with a single power-law differential slope β = 1.9 ± 0.1, comparable with the average XLF of other elliptical galaxies (Kim and Fabbiano, 2004). In NGC 4278 we find that the XLF follows this single power-law down to the 90% completeness limit of 3 x $10^{37}$ erg $s^{-1}$, and that this trend extends down to $10^{37}$ erg $s^{-1}$ in the bias-corrected XLF. In 3379, there could be a break at ~4 x $10^{37}$ erg $s^{-1}$ (although the discrepancy from a single power law fit is only a 1.6σ effect), but the XLF becomes steep again below 2 x $10^{37}$ erg $s^{-1}$ and follows this shape at least down to ~5 × $10^{36}$ erg $s^{-1}$. We can exclude that this steep power-law at the low luminosities is the result of some problem with our bias correction, because it is directly observed in the complete portion of the merged XLF (see Fig. 3). In summary, our results show that there is no universal flattening of the XLF at the low luminosities.

We note that although a flattening was reported in the XLF of NGC 5128, there are several concerns that may affect these results, although the authors (Voss & Gilfanov 2005) tried to take them into account. NGC 5128 is far from being a normal elliptical, since its X-ray source population may have been rejuvenated by a merger event, and mixing with a population of HMXBs (which is known to have a shallower XLF; e.g., Grimm, Gilfanov & Sunyaev 2003) is possible; HMXB contamination may also affect the XLFs of the Milky Way and M31. Also, the X-ray emission of NGC 5128 is complex, with various sub-structures in the hot ISM as well as strong dust lanes, optical filaments, radio/X-ray jets, etc. (e.g., Karovska et al. 2002), potentially increasing the difficulty of obtaining a bias-free low luminosity XLF. Moreover, the large angular size ($D_{25}$ = 26′) of NGC 5128 makes it hard to effectively remove contamination by foreground and background X-ray sources, which, given the depth of the exposure, account for about half of the detected sources (Voss & Gilfanov 2005). We note that at the distance of NGC 5128 (3.5 Mpc), the break in the cosmic X-ray background log(N)-log(S) relation at $f_X$ ~ 2.5 x $10^{-14}$ erg $s^{-1}$ $cm^{-2}$ (Kim, M. et al, 2006) happens to correspond to $L_X$ ~ 4 x $10^{37}$ erg $s^{-1}$, close to the reported break.

Many models predict that the LMXB XLF should flatten at the low luminosities, although the exact break luminosity is not well determined (see review Fabbiano 2006). Short-period binaries are expected to have a XLF break at this luminosity, based on the 'outburst peak luminosity – orbital period' correlation (King & Ritter 1998). Flattening of the XLF is found in the population synthesis of field LMXBs of Pfahl, Rappaport & Podsiadlowski (2003, their fig. 3), if irradiation of the donor star from the X-ray emission of the compact companion is considered in the model. The model of Bildsten & Deloye (2004) of formation of ultra-compact LMXBs in GCs also predicts a flattening at the lower luminosities ($L_X$ ~$10^{37}$ erg $s^{-1}$), where these sources would become transient and therefore disappear from the XLF, but this break may also occur at much lower luminosities if X-ray heating stabilize the accretion disk (see also Deloye & Bildsten 2003). In this light, we note that if the possible XLF bump, or break, at $L_X$ ~ 4 x $10^{37}$ erg $s^{-1}$ in NGC 3379 is real, this (and the absence of the XLF break in NGC 4278) could suggest a higher fraction of ultra-compact, X-ray-heating stabilized, low-luminosity sources in NGC 4278, consistent with the higher GC specific frequency ($S_N$) of this galaxy (see Table 2). More recently, Postnov & Kuranov (2005) argue that a low-luminosity break would be related to different accretion driving mechanisms in LMXBs:

in the lower luminosity sources ($L_X$ < a few x $10^{37}$ erg s$^{-1}$) accretion may be driven by angular momentum removal by gravitational waves, while at higher luminosities (a few x $10^{37}$ < $L_X$ < 5 x $10^{38}$ erg s$^{-1}$) magnetic stellar winds may be responsible. Optical identification of X-ray sources with GCs and an estimate of the transient fraction at different luminosities (which we will attempt once the full body of observations on NGC3379 and NGC4278 will be completed), may help discriminating among possible scenarios.

It can't be excluded that there are differences in the XLFs of different LMXB populations, possibly reflecting the origin and evolution of these LMXBs. Even ignoring the possible bump in NGC 3379, while the XLF slopes of our two galaxies are similar, their amplitudes (i.e. total LMXB X-ray luminosity or total number of LMXBs) differ by a factor of 4 (Table 2). Because the optical luminosities are similar (Table 1), this discrepancy is not related to the total stellar mass, which has been suggested as the main driver of the normalization of the LMXB XLF (Gilfanov 2004). This discrepancy is of the order of the maximum amount of the scatter observed in the relation between the X-ray luminosity of LMXBs and the optical luminosity in 14 early type galaxies by Kim and Fabbiano (2004). In our two galaxies, the specific frequency of GC ($S_N$) is very different (column 8 in Table 1). The LMXB population of NGC3379 may be dominated by field binaries, while that of NGC4278 may have originated prevalently in globular clusters. The ratios $L_X$ (LMXB)/$L_K$ of these two galaxies and their $S_N$ are consistent with the $L_X$ (LMXB)/$L_K$ - $S_N$ relation made with a larger sample in Kim & Fabbiano (2004; see their Figure 8), because N4278 has a considerably higher $S_N$ than N3379. The LMXB – GC connection and the comparison between field and GC LMXBs will be explored in a future paper.

5. CONCLUSIONS

We have derived and analyzed the deepest XLF of LMXBs in two normal old stellar population elliptical galaxies: NGC 3379 and NGC 4278. The XLF of NGC 4278 reaches down to 1 × $10^{37}$ ergs s$^{-1}$ and follows closely a straight, unbroken, power-law of differential slope 1.9 ± 0.1, with high statistical significance. This result shows that the 'universal' break of the LMXB XLF at ~5 × $10^{37}$ ergs s$^{-1}$, suggested by Gilfanov (2004), is not universal. The XLF of NGC 3379 is also well fitted by a similar power law, although there is a possible (1.6 σ) excess above this power law at ~ 4 × $10^{37}$ ergs s$^{-1}$. Even if this excess is real, it may not represent a break, because the power-law resumes at the lower luminosities, below ~2 × $10^{37}$ ergs s$^{-1}$.

Comparison of the normalizations of the XLFs (i.e. the integrated LMXB luminosities of the galaxies), shows that for the same integrated stellar luminosity, NGC 4278 is ~4 times overluminous in X-rays. This results confirms the conclusion of Kim & Fabbiano (2004; see also White et al. 2002) that although stellar luminosity / mass is an important driver of the XLF normalization, as suggested by Gilfanov (2004), it is not the sole driver. The specific frequency of globular clusters in a galaxy is also important: while NGC 3379 has few GCs, NGC 4278 is GC rich.

This paper is the first report on the deep XLF study of LMXB populations that is being currently performed with a legacy Chandra observing campaign. Future work will push these XLFs to even lower luminosities and will explore separately the contributions of field and GC LMXBs to the XLF.

**ACKNOWLEDGEMENTS**

This work was supported by the Chandra GO grant G06-7079A (PI Fabbiano) and subcontract GO6-7079B (PI Kalogera). D.-W. Kim acknowledges support from NASA contract NAS8-39073 (CXC); A. Zezas acknowledges support from NASA LTSA grant NAG5-13056. The data analysis was supported by the CXC CIAO software and CALDB.

**Figure Captions**

Figure 1. Merged *Chandra* images of (a) NGC 3379 and (b) NGC 4278. The large green ellipse indicates the optical galaxy size at the $25^{th}$ magnitude and the small circles indicate detected point sources.

Figure 2. X-ray luminosity function of LMXBs determined with the merged observations of (a) NGC 3379 and (b) NGC 4278. Filled squares with error bars indicate the bias-corrected XLF and green circles indicate the raw data. The best fit single power law model is plotted by the histogram. The dotted line indicates the expected number of cosmic X-ray background sources determined with the ChaMP + CDF data (Kim, E. et al. 2006).

Figure 3. XLFs determined with two individual observations are compared with that of the merged observation. The black and green histograms indicate the corrected and uncorrected XLF. The vertical bar indicates a 90% completeness limit (column 9 in Table 2). The XLFs determined in two individual observations are shifted down vertically by $\Delta \log(N) = -1.0$ (the shallower one) and $-0.5$ (the deeper one). The best-fit single power-law model (dashed blue diagonal lines) determined with the merged observations is plotted with observed XLFs (for both individual and merged).

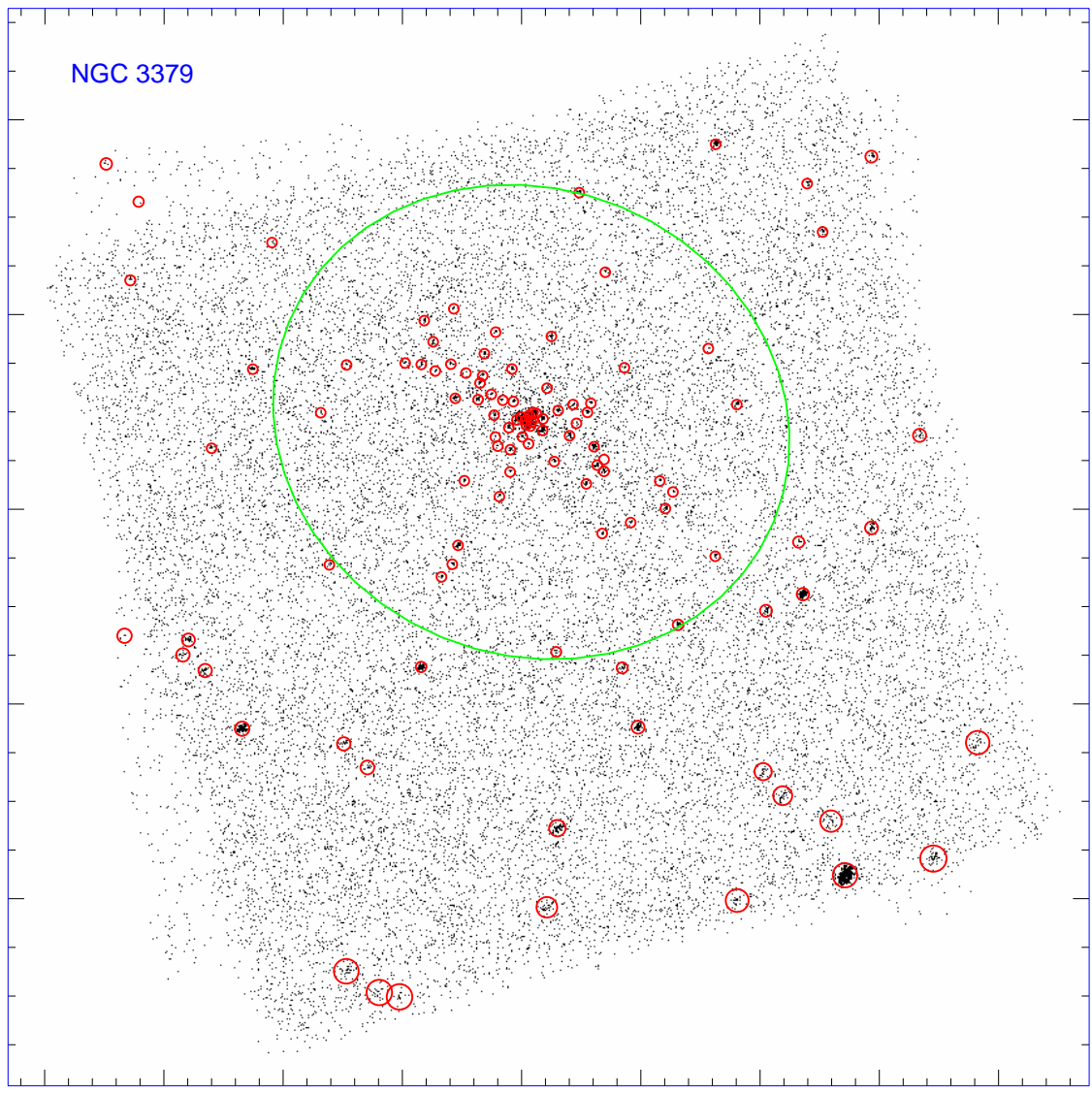

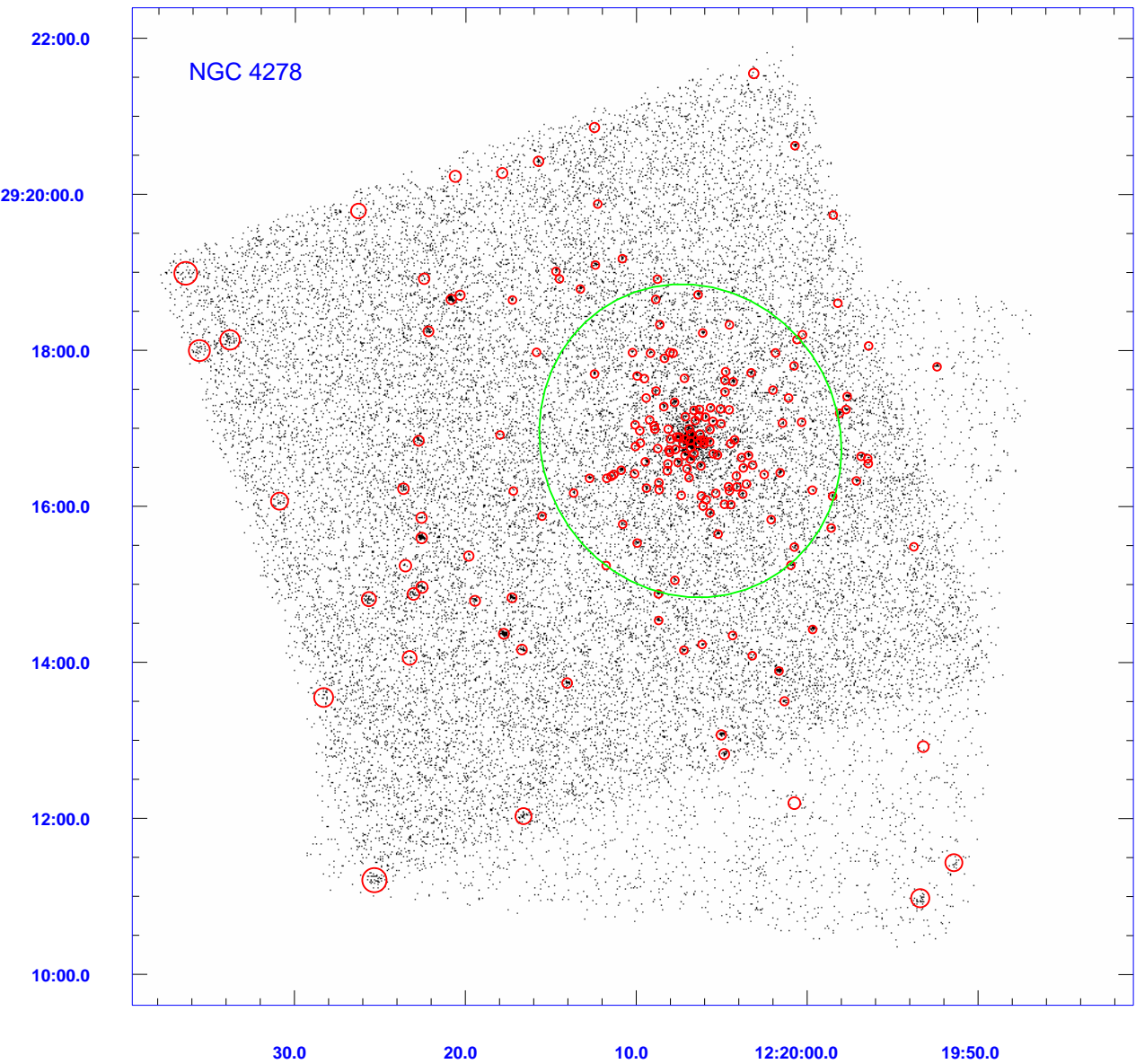

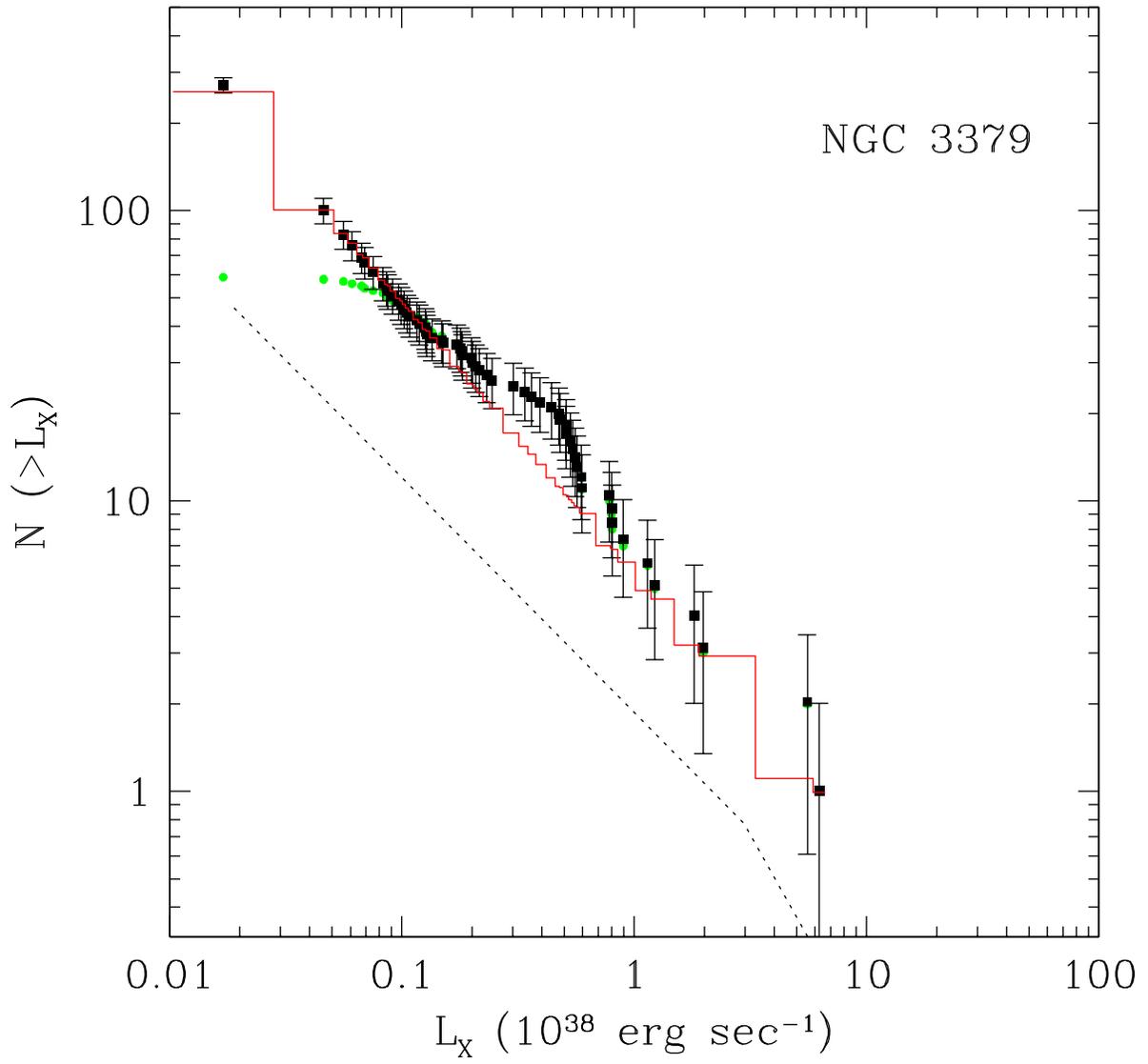

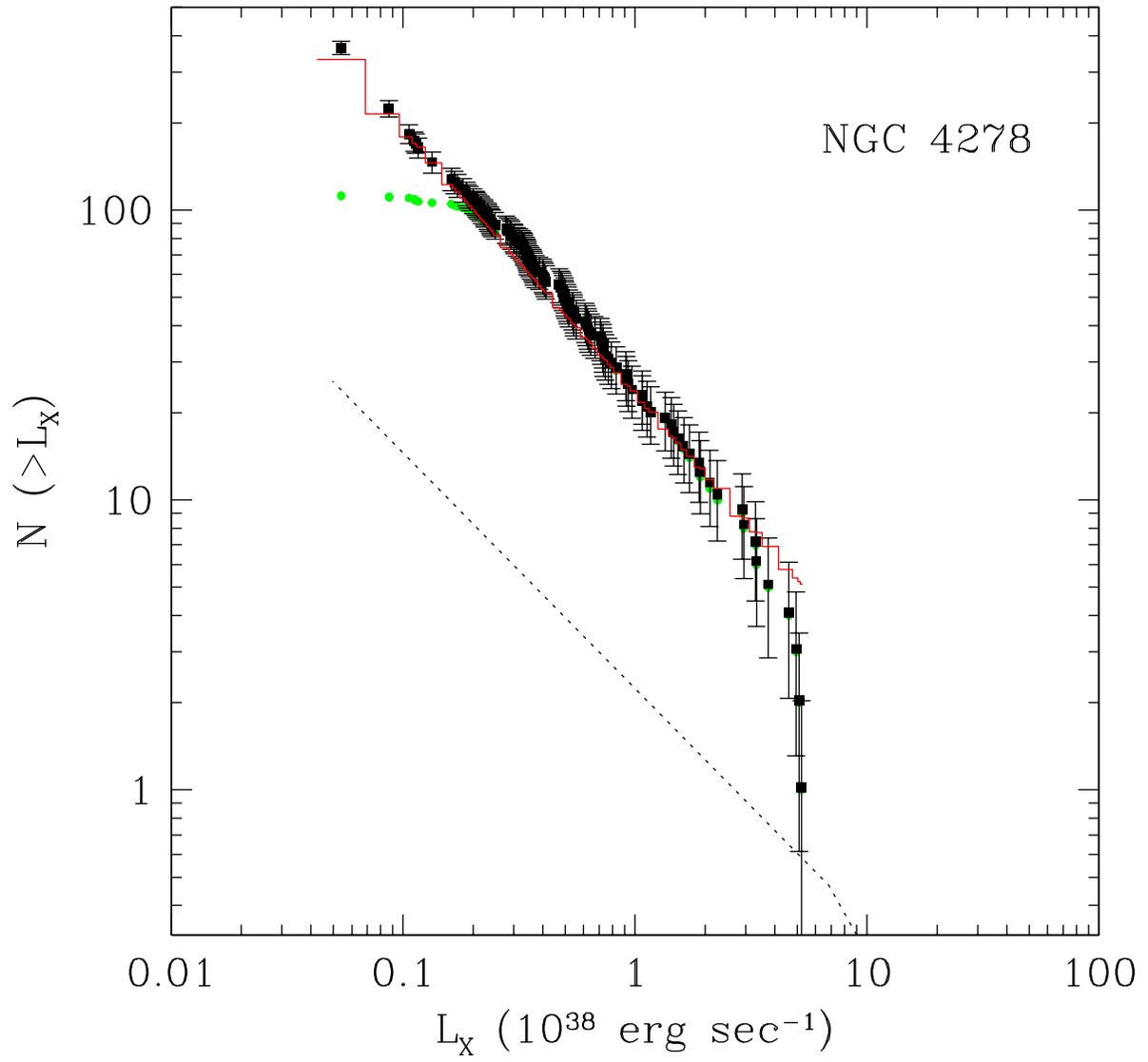

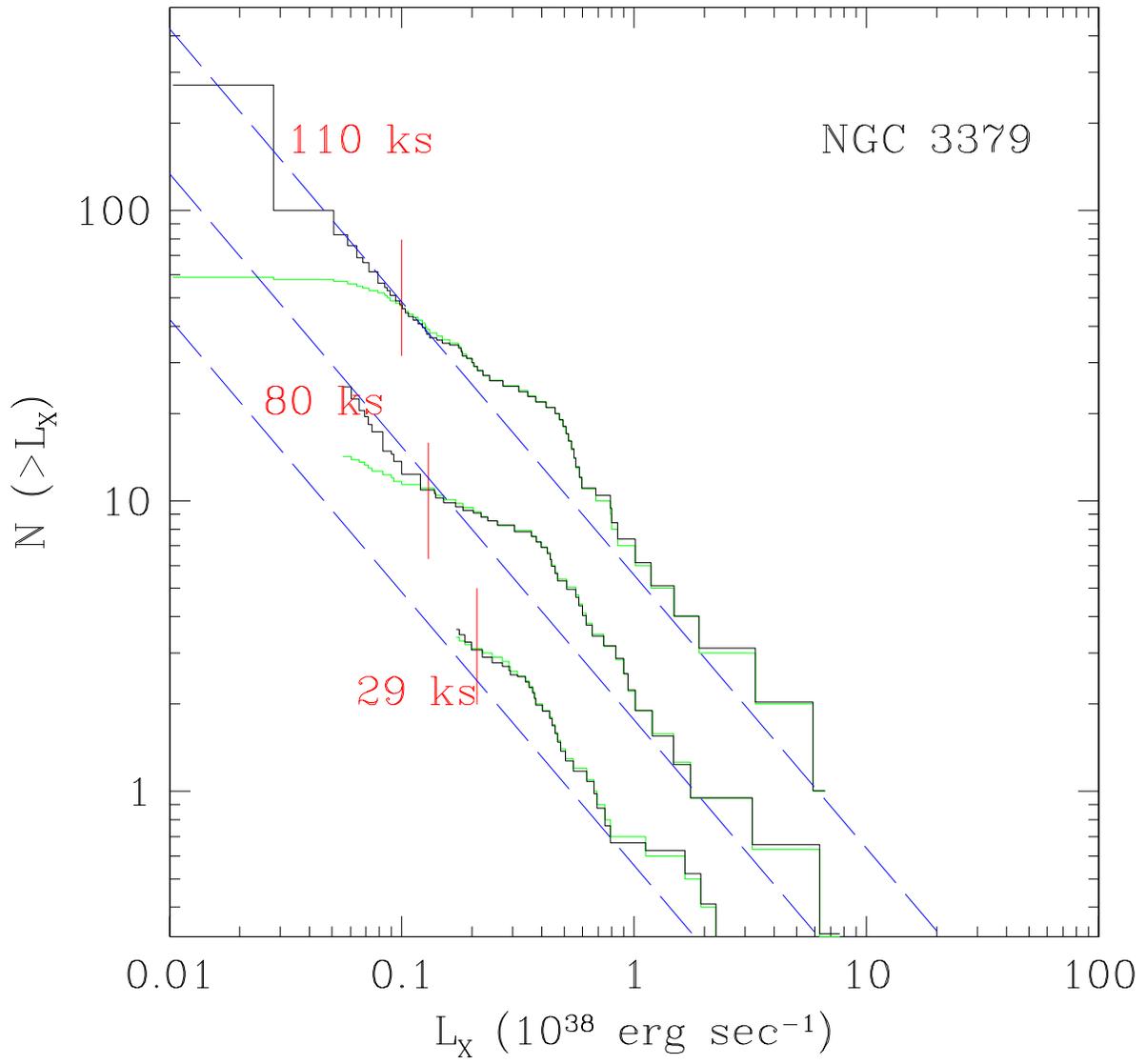

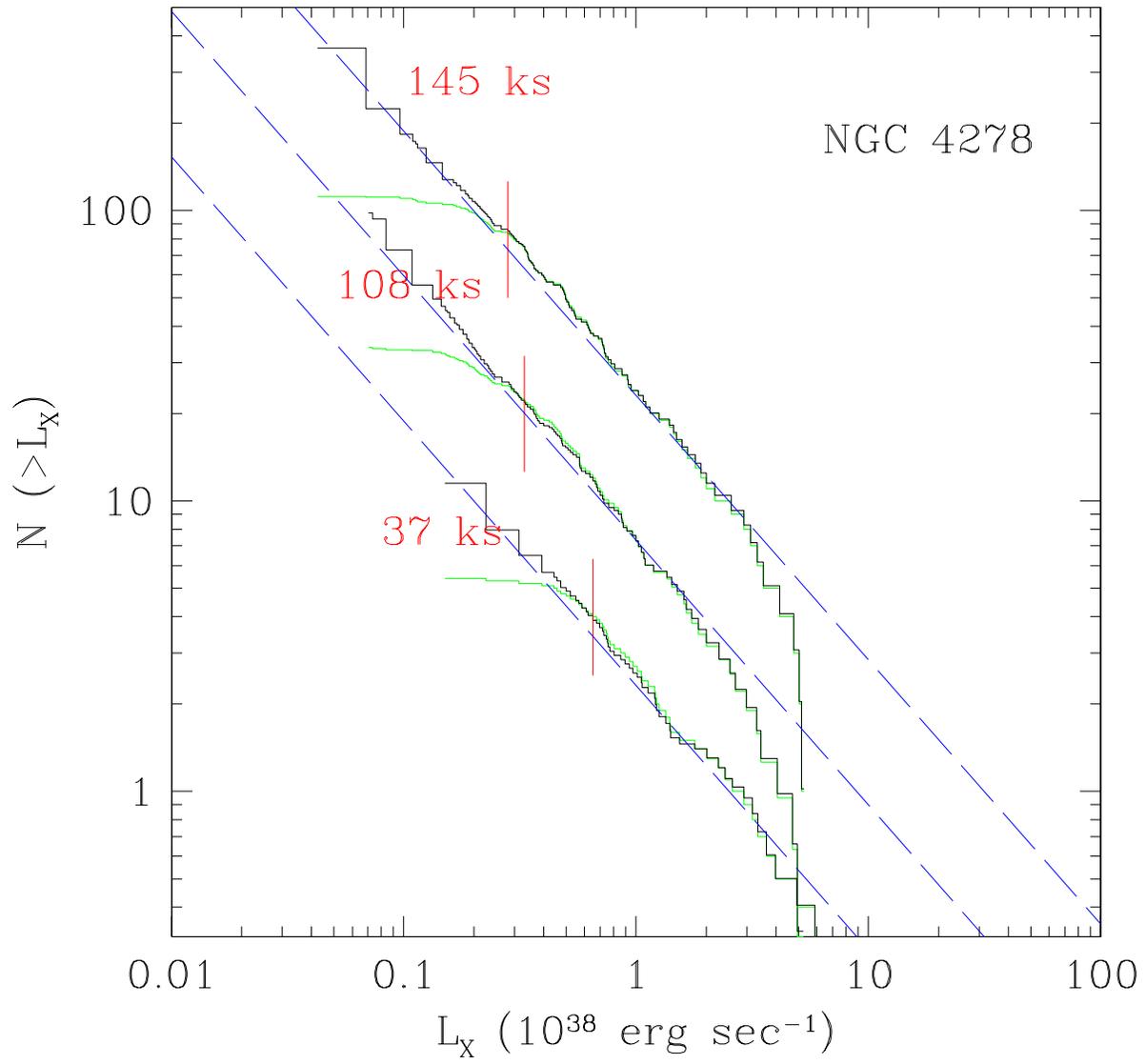